\documentclass{article}
\usepackage{ijcai22}

\usepackage{times}
\usepackage{soul}
\usepackage{url}
\usepackage[hidelinks]{hyperref}
\usepackage[utf8]{inputenc}
\usepackage[small]{caption}
\usepackage{graphicx}
\usepackage{amsmath}
\usepackage{amsthm}
\usepackage{booktabs}
\usepackage{algorithm}
\usepackage{algorithmic}
\usepackage[switch]{lineno}
\usepackage{tikz}
\usepackage{makecell}
\usepackage{framed}
\usepackage{mdframed}

\usepackage{footnote}
\usepackage{amsfonts}
\newtheorem{definition}[]{Definition}
\newtheorem{theorem}{Theorem}
\newtheorem{proposition}{Proposition}
\newtheorem{lemma}{Lemma}
\usepackage{chngcntr}
\usepackage{apptools}

\newtheorem{example}{Example}

\title{Task Allocation on Networks with Execution Uncertainty}

\author{
Xiuzhen Zhang
\and
Yao Zhang\And
Dengji Zhao
\affiliations
School of Information Science and Technology, ShanghaiTech University
\emails
\{zhangxzh1, zhangyao1, zhaodj\}@shanghaitech.edu.cn
}

\begin{document}
\setlength{\aboverulesep}{0pt}
\setlength{\belowrulesep}{0pt}

\maketitle

\begin{abstract}
We study a single task allocation problem where each worker connects to some other workers to form a network and the task requester only connects to some of the workers. The goal is to design an allocation mechanism such that each worker is incentivized to invite her neighbours to join the allocation, although they are competing for the task. Moreover, the performance of each worker is uncertain, which is modelled as the quality level of her task execution. The literature has proposed solutions to tackle the uncertainty problem by paying them after verifying their execution. Here, we extend the problem to the network setting. The challenge is that the requester relies on the workers to invite each other to find the best worker, and the performance of each worker is also unknown to the task requester. In this paper, we propose a new mechanism to solve the two challenges at the same time. The mechanism guarantees that inviting more workers and reporting/performing according to her true ability is a dominant strategy for each worker. We believe that the new solution can be widely applied in the digital economy powered by social connections such as crowdsourcing and contests.
\end{abstract}

\section{Introduction}
Task allocation is an important part of real-world applications such as crowdsourcing~\cite{wu2017efficient,goel2014allocating} and market supply~\cite{dash2007market}. A common goal of task allocation is to find suitable workers to achieve a good performance at a low cost.
Previous studies have made great progress in finding the best allocations under cases with a fixed number of workers. For example, the task requester seeks suitable workers in third-party platforms (e.g., Amazon Mechanical Turk) or holds a contest with attractive rewards~\cite{chawla2019optimal}. Yet, such cases are less scalable due to the relatively fixed number of participants. Generally, we hope to involve more workers so that the task requester is capable of finding more suitable workers. Also, nowadays, people are connected with others via social networks. Therefore, a straightforward approach is to make full use of their connections such that we can involve more workers. The challenge remains such as workers are competitors for the task and they are unwilling to provide their connections.

More precisely, we consider a single-task allocation problem where the task is allocated to a single agent and the task performance of an agent is measured by the finished quality. Each agent has a cost to perform the task. Before conducting the tasks, agents are uncertain about their actual performance and only know their \textit{probability distributions over the quality levels}, which is also known as the execution uncertainty. Then, another challenge arises to the task requester for the robustness of the task allocation to the execution uncertainty.

We propose the \textit{PEV-based Diffusion Mechanism} to handle the challenges one by one. Firstly, to solve the issue of agents unwilling to invite others, the proposed mechanism tries to reward them such that each agent will invite all her neighbours to maximize her utility. Then, the task requester is able to reach as many agents as possible. Secondly, to allow for the execution uncertainty, the proposed mechanism gives agents payoffs based on their actual performance, which guarantees that agents will not misreport their abilities and the task requester will not have a deficit in expectation. More importantly, previous studies focused on the uncertain successful performance, introducing the probability of success (PoS) to describe the probability of an agent successfully completing the task~\cite{porter2008fault,ramchurn2009trust}, e.g., 70 \% to fail and 30 \% to finish the task. Yet such a metric is weak to accurately describe agents' abilities. In our setting, we define the probability of quality (PoQ), which represents the probability distribution on completion qualities, e.g., 30 \% to finish with a good quality, 20 \% to finish with a low quality, and 50 \% to totally fail.

To sum up, the goal of this paper is to design a mechanism to incentivize agents to invite all their neighbours, report their PoQs and costs to perform the task. The mechanism should also guarantee that the task requester will not suffer a loss compared to the case where the participating agents are all the task requester's neighbours. We first consider the case where the task requester has no requirements for agents' abilities such that each agent can perform the task in the same quality with a probability of one. We show that the Information Diffusion Mechanism~\cite{li2017mechanism,li2020incentive} can be applied in such a case. However, when the task requester is sensitive to agents' completion qualities and uncertain about agents' performance, the Information Diffusion Mechanism fails in incentive compatibility. To solve this issue, we then propose the PEV-based Diffusion Mechanism to meet these requirements.





 


\subsection{Related Work}
The social network is an effective medium to get access to more potential agents. Mechanism design in social networks has been widely utilized in auctions~\cite{li2020incentive}, answer querying~\cite{tang2011reflecting}, social advertising~\cite{li2012diffusion} and influence maximization~\cite{DBLP:conf/ecai/ShiZSWZ20}.
In this paper, we are inspired by the idea of the Information Diffusion Mechanism~\cite{li2020incentive,li2017mechanism}, which is proposed to increase the seller' revenue in auctions via social networks. The Information Diffusion Mechanism designs the payoffs based on their contributions to find the buyer with the highest bid. Though there has also been work studying task allocation problems on social networks~\cite{jiang2012task,de2012multiagent},
diffusion incentives and strategic actions to hide connections were not taken into their consideration, while these are the main concerns in our setting. 

In traditional task allocation problems, the performance and cost that each agent achieves are private information. To achieve truthfully reporting, the task scheduling mechanisms with verification were first proposed to take both agents' declarations and their actual performance into consideration~\cite{nisan2001algorithmic,DBLP:conf/aaai/ConitzerV14}.
Later, since there exist cases where agents may fail to reach the same performance as they declared in real-world applications,
the execution uncertainty is considered in mechanism design problems. To describe the execution uncertainty, 
probability of success (PoS) is introduced to describe the probability of an agent successfully completing the tasks~\cite{porter2008fault,ramchurn2009trust,zhao2016fault}. 

The remainder of this paper is organized as follows. Section~\ref{sec:model} describes the model of the single-task allocation problem in social networks and introduces desirable properties. Section~\ref{sec:mechanism} presents the Information Diffusion Mechanism in the setting without the execution uncertainty and shows the failure of its application in the general setting. Following that, we propose our PEV-based Diffusion Mechanism and prove its remarkable performance. We conclude in Section~\ref{sec:conclude}.


\section{The Model}\label{sec:model}
Consider a social network represented by a graph $G=(V,E)$, where $V = \{s\} \cup N$ is the node set and $E$ is the edge set. The task requester $s$ has a single task to be performed and $N = \{1,2,\cdots,n\}$ is the set of all other agents in the network. Each edge $(i,j) \in E$ indicates that agent $i$ can directly communicate with agent $j$. For $i \in V$, let $r_i = \{j\in V\mid (i,j)\in E\}$ be the neighbour set of $i$. Given the task to be performed, let $Q \subset \mathbb{R}^+ \cup \{0\}$ be the set of all possible completion qualities. Let the discrete random variable $Q_i$ be the completion quality of agent $i$ and $q_i \in Q$ denote a realization of $Q_i$. Let $f_i$ be the probability density function of $Q_i$, i.e., $P(Q_i = q_i) = f_i(q_i)$. The probability distribution $f_i$ is called agent $i$'s probability of quality (PoQ). There is also a fixed cost $c_i \geq 0$ for $i$ to perform the task. Define $\theta_i = (f_i,c_i,r_i)$ as agent $i$'s type, which is only known to her. Let $\Theta_i$ be the type space of agent $i$ and $\theta = (\theta_1,\cdots,\theta_n)$ be the type profile of all agents.


Under the above setting, the goal of the task requester is to assign the task to an agent who can perform it with a high quality and a low cost. Initially, only the task requester's neighbours $r_s$ know the task and they may not be the best worker for the task. Hence, the task requester needs a mechanism to attract more participants, which is done by incentivizing agents to diffuse the task information to all their neighbours. Thus, each agent's action consists of reporting her PoQ, her cost to perform the task and inviting her neighbours, i.e., reporting her type. For agent $i \in N$, let $\theta'_i = (f_i',c_i',r_i')$ be her report, where $f_i'$ is a probability distribution over $Q$, $r'_i\subseteq r_i$ and $c_i' \geq 0$. Let $\theta'$ $=$ $(\theta'_1,\theta'_2,\cdots,\theta'_n)$ be a report profile of all agents in $N$.
Denote the graph constructed from $\theta'$ by $G(\theta') = \left(V, E(\theta') \right)$, where $E(\theta') =  r_s \cup \{(i,j) \mid i \in N, j \in r_i'\}$.
Let $I(\theta')$ be the set of all participants under $\theta'$, and $i \in I(\theta')$ holds if and only if there exists a path from $s$ to $i$ in the graph $G(\theta')$. Let $\Theta$ be the space of all possible type profiles. 


Generally speaking, the mechanism consists of two steps. The task requester first announces a contract including a task allocation policy and a payoff policy and then assigns the task to an agent according to their declarations (in our setting, we only consider the case where the task is assigned to at most one agent). After the task requester verifies the completion quality, she will give payoffs to agents according to the announced contract. We call such a mechanism the verified contract mechanism and provide the formal definition below.

\begin{definition}[Verified Contract Mechanism]
A \textbf{verified contract mechanism} is defined by $\mathcal{M} = (\pi, p)$, where $\pi: \Theta \rightarrow \{0,1\}^{N}$ and $p: \Theta\times Q \rightarrow \mathbb{R}^{N}$ are the allocation and payoff policies respectively. Given a set of agents $N$ and all agents' report profile $\theta' \in \Theta$, set $\pi_i(\theta') = 0$ and $p_i(\theta',q) = 0$ for all $i \notin I(\theta')$, $q$ is the quality that the task requester receives.
\end{definition}

Given a verified contract mechanism and a report profile $\theta'$, let $\pi(\theta')$ be the allocation of the verified contract mechanism. $\pi_i(\theta')$ $=$ $1$ means that the task is allocated to agent $i$, otherwise she will not perform the task. Since the task can only be assigned to at most one agent, we have $\sum_{i\in N} \pi_i(\theta') \leq 1$. The actual completion quality under the allocation $\pi$ is drawn from the true PoQ of the selected agent, denoted by $q_{\pi}$.
Then $p_i(\theta',q_{\pi})$ is the payoff to agent $i$ given from the task requester $s$. We assume that the utilities of the task requester and the agents are quasi-linear, i.e., $u_s(\theta', q_{\pi}) = q_{\pi} - \sum_{i\in N} p_i(\theta', q_{\pi})$ and $u_i(\pi(\theta'), p(\theta', q_\pi)) = p_i(\theta', q_{\pi}) - \pi_i(\theta') c_i$ for all $i\in N$
(In reality, we can map the evaluation of the quality into the same measure of payoffs and costs). In the following, we define several properties concerned in our model. The first one is the efficiency of the mechanism in terms of the expected social welfare. 

\begin{definition}[Efficiency]
A verified contract mechanism is \textbf{efficient} if for all $\theta' \in \Theta$, 
\[ \pi(\theta') \in \arg\max\nolimits_{\pi' \in \Pi} \mathbb{E}_{f_{\pi'}'} \left[q_{\pi'} - \sum\nolimits_{i \in N }\pi'_i c_i' \right] \]
where $\Pi$ is the space of all feasible allocations, $f_{\pi'}'$ is the PoQ reported by the selected agent under $\pi'$ (may not be her true PoQ) and $\mathbb{E}_f[X]$ is the expected value of $X$ taken over $f$.
\end{definition}

In other words, the social welfare equals the completion quality minus the performing cost. An efficient mechanism maximizes the expected social welfare of all agents in the network. Another property is called incentive compatibility, which requires that for each agent $i$ participating in the mechanism, reporting her type $\theta_i$ truthfully is a dominant strategy.

\begin{definition}[Incentive Compatibility]
A verified contract mechanism is \textbf{incentive compatible} (IC) if for all $i \in N$ and $\theta_i \in \Theta_i$, for all $\theta'_{-i} \in \Theta_{-i}$,
\[ \theta_i \in \arg\max_{\theta'_i \in \Theta_i} \mathbb{E}_{f_i}\left[ u_i(\pi((\theta_i', \theta_{-i}')),p((\theta_i', \theta_{-i}'), q_{\pi((\theta_i', \theta_{-i}'))})) \right],\]
where $\theta_{-i}'$ is the report profile of all agents without $i$, $\Theta_{-i}$ is the space of all possible $\theta_{-i}'$.
\end{definition}

Intuitively, incentive compatibility ensures that agents are incentivized to truthfully reveal their abilities and costs to finish the task, and propagate the task information to all their neighbours. The next property ensures that the expected utility of an agent is non-negative when she truthfully reports.

\begin{definition}[Individual Rationality]
A verified contract mechanism is \textbf{individually rational} (IR) if for all $i \in N$ and $\theta_i \in \Theta_i$, for all $\theta'_{-i} \in \Theta_{-i}$,
\[\mathbb{E}_{f_i}\left[ u_i(\pi((\theta_i, \theta_{-i}')),p((\theta_i, \theta_{-i}'), q_{\pi((\theta_i, \theta_{-i}'))})) \right] \geq 0.\]
\end{definition}

The property of individual rationality ensures that agents will not lose to participate in the mechanism. The last desirable property is that the task requester should not suffer a deficit in the task allocation. More precisely, the task requester should consistently achieve a non-negative expected utility, which guarantees the applicability of the mechanism.

\begin{definition}[Weakly Budget Balance]
A verified contract mechanism is \textbf{weakly budget balanced} (WBB) if for all $\theta' \in \Theta$,
\[\mathbb{E}_{f_{\pi(\theta')}} \left[ u_s(\theta',q_{\pi(\theta')}) \right] \geq 0,\]
where $f_{\pi(\theta')}$ is the true PoQ of the selected agent under $\theta'$.
\end{definition}

\section{The Mechanism}\label{sec:mechanism}
In this section, we first consider a setting without the execution uncertainty. When no special skills are required to finish the task, all agents perform the given task with the same quality though their costs can differ. We show that the Information Diffusion Mechanism is effective in this setting. However, when the task requires special skills, agents may perform the task with different qualities. We find that the Information Diffusion Mechanism fails in incentive compatibility in this setting. Accordingly, we propose our mechanism, called the PEV-based Diffusion Mechanism, to adapt to both settings. In the PEV-based Diffusion Mechanism, agents are incentivized to diffuse the task information and truthfully report their abilities and costs. The details are illustrated below.

\subsection{Without Execution Uncertainty}
We first consider the setting where the task performance doesn't require special skills. The task requester gets the same quality no matter which agent the task is assigned to. Then, it is without loss of generality to assume that $Q = \{q\}$ and $f_i(q) = 1$ for all $i \in N$. However, agents may need different costs to perform the given task (e.g., time). 
Thus, to maximize both the task requester's utility and the social welfare, a mechanism should allocate the task to an agent who can perform the task with the least cost.

\begin{figure}[!htbp]
    \centering
    \includegraphics[scale = 0.42]{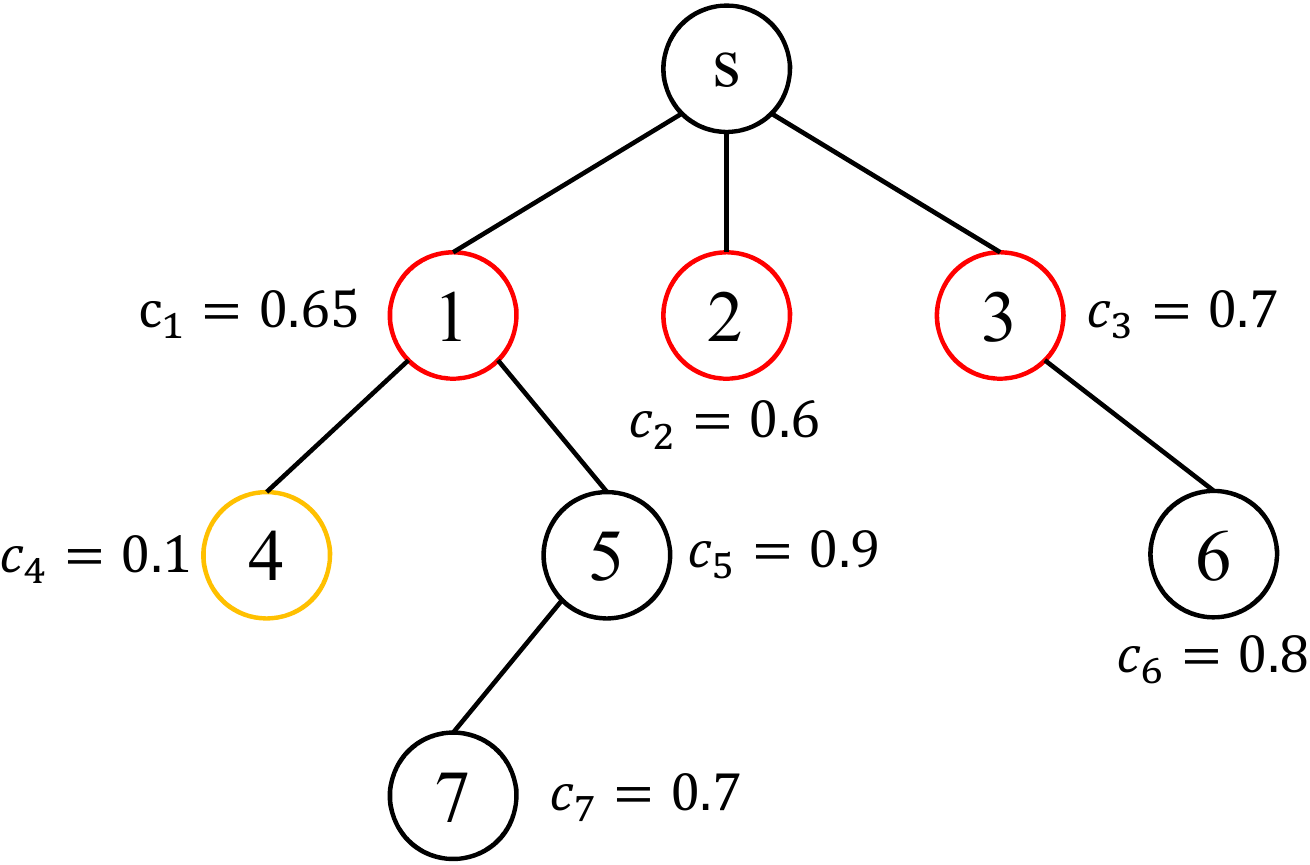}
    \caption{\iffalse A social network contains $7$ possible workers.\fi Given a task, agents' costs to perform the task are shown in the figure. Each agent $i$ can perform the task with $f_i(q) = 1$ and $q= 1$. Without diffusion, the task requester $s$ can only know agent $1$, $2$ and $3$, and agent $2$ is chosen. If agents invite all their neighbours, $s$ can know the whole network and then chooses the agent $4$.}
    \label{fig:ex1}
\end{figure}

Without considering the social network diffusion, all participants are the task requester's direct neighbours. 
The VCG (Vickrey–Clarke–Groves) mechanism is proposed to incentivize truthfully reporting in auctions. It can be applied here if the goal is changed to select an agent who can perform the task with the least cost rather than selecting the agent with the highest bid. Applying the VCG mechanism, the task requester selects the agent who performs with the least cost among her neighbours, and the agent's payoff equals the decrease in others' utilities due to her participation. 
Yet the agent may not be the best worker in the social network. 
For example, let's consider a network in Figure~\ref{fig:ex1}. 
Without diffusion, the task requester will choose agent $2$ while agent $4$ can perform the task with the least cost among 7 possible workers. To maximize social welfare, the goal here is to design a mechanism where all agents are incentivized to truthfully report their costs and invite all their neighbours.

When considering network diffusion, the VCG mechanism cannot be applied since it will cause deficits for the task requester. An intuitive example is shown in Example~\ref{vcgex}.

\begin{example}\label{vcgex}
Consider the social network shown in Figure~\ref{fig:ex1}, agent $4$ is chosen as the agent to perform the task. Let $w$ be the social welfare under this allocation and $w_{i}$ be the maximal social welfare without $i$'s participation. Then we have $w_{1} = q - c_2 = 0.4$, $w_{4} = q - c_2 = 0.4$ and $w = q - c_4 = 0.9$. Therefore, the payoff to agent $1$ is $p_1 = q - c_4 - w_{1} = 0.5$, the payoff to agent $4$ is $p_4 = q - w_{4} = 0.6$ and the payoffs to all other agents are $0$. Then, the utility of the task requester is $u_s = q - p_1 - p_4 = -0.1$.
\end{example}

To tackle this issue in this setting, we can apply the Information Diffusion Mechanism~\cite{li2017mechanism}. The Information Diffusion Mechanism is proposed to incentivize the auction information diffusion to sell a single item in a social network. Similarly, the goal is also changed to select the agent with the least cost. 
Before giving the mechanism, we first define \textit{critical agents} as follows.
\begin{definition}
Given a report profile $\theta'$, for agent $i,j \in I(\theta')$, $j$ is one of $i$'s \textbf{critical agent} if $j$ exists in all simple paths from the task requester $s$ to agent $i$ in the graph $G(\theta')$. The set of $i$'s all critical agents is denoted by \textbf{$ct_i(\theta')$} and the sequence order of agents in $ct_i(\theta')$ is denoted by $(s, j_1,\cdots,j_m)$, where $j_m = i$ and $k<k'$ if and only if $j_k \in ct_{j_{k'}}$.
\end{definition}
 Then, with the definition above, we apply the Information Diffusion Mechanism in our setting as the following.
\begin{framed}
 \noindent\textbf{Information Diffusion Mechanism}
 
 \noindent\rule{\textwidth}{0.5pt}
 
 \noindent\textsc{Input}: a report profile $\theta'$.
 
 \noindent\rule{\textwidth}{0.5pt}
 1. Choose $j \in \arg\max\nolimits_{i \in I(\theta') } \{q-c'_i \}$
     with random tie breaking.
     
\noindent    
2. Set $\pi_i = 0, p_i = 0$ for all $i \notin I(\theta')$ and all $i \notin ct_j(\theta')$.

\noindent
3. Compute $w_i = \sum_{k \neq i}\pi_k' (q-c_k')$ for each $i \in ct_j(\theta')$, where $\pi' = \left\{ \pi_i'( \hat{\theta} ) = \mathbb{I} \left( i \in \arg\max\limits_{k \in I(\hat{\theta} )}q- c_k' \right) \right\}_{ i \in N}$ and $\hat{\theta} = \left(nil,\theta_{-i}'\right)$.

\noindent
4. Let the sequence order of agents in $ct_j(\theta')$ be $ (s,i_1,\cdots,i_m)$, where $i_m = j$. For $k = 1:m-1$,
     \[\pi_{i_k} = \begin{cases} 1 & \text{if } q- c_{i_k}' = w_{i_{k+1}} \\ & \text{ and } \sum_{l=1}^{k-1} \pi_{i_l} = 0 \\ 0 & \text{otherwise}   \end{cases}\]
     and $\pi_j = 1$ if $\sum_{k=1}^{m-1}\pi_{i_k} = 0$.
     
\noindent
5. Suppose $i_t$ be the chosen agent, i.e., $\pi_{i_t} = 1, 1\leq t \leq m$. Then, the payoff of each agent $i_k \in ct_j(\theta')$ will be 
     \[ p_{i_k} = \begin{cases} w_{i_{k+1}} - w_{i_k} & k < t \\ q- w_{i_t}  & k = t \\ 0 & k>t.  \end{cases} \]
 
 \noindent\rule{\textwidth}{0.5pt}
 
 \noindent\textsc{Output}: the allocation $\pi$ and the payoff $p$.
\end{framed}

Intuitively, the Information Diffusion Mechanism allocates the task to one agent in the sequence of $j$'s critical agents. More precisely, the mechanism will choose the first agent with the least cost when the next agent in the sequence did not participate\footnote{Assume that there exists at least one agent whose cost to perform the task is less than $q$, otherwise we can add a dummy agent $d$ with $c_d=q$, and the payoff to the dummy agent is always 0 to ensure the social welfare will be non-negative.}. The payoff of the chosen agent equals the least cost reported by other agents who can still participate without her invitation. The payoff of each agent before the chosen agent in the sequence is determined by the difference between the maximum social welfare without her next agent's participation and that without her participation. 

\begin{theorem}[\citeauthor{li2017mechanism}~\cite{li2017mechanism}]
The Information Diffusion Mechanism is individually rational, incentive compatible and weakly budget balanced.
\end{theorem}
\noindent
\textbf{Failure in the Quality-Aware Setting.}
We then consider another setting without execution uncertainty. In this setting, the task needs special skills to be performed, and then agents may perform the task with different qualities, i.e., $f_i(q_i) = 1$ for $q_i\in Q$ and $i \in N$.
Each agent reports her PoQ as $f_i'$, where $f_i'(q_i') = 1$ with $q_i' \in Q$.
Then, the task requester also focuses on the qualities that agents will perform the given task with, and we call this setting a quality-aware setting. In this setting, the Information Diffusion Mechanism will try to choose an agent with high-quality performance and a low cost, which then leads to the application failure. The failure lies in no guarantee for incentive compatibility since the payoff to the selected agent is always related to her reported ability. The mechanism will also cause a deficit to the task requester since she may get low-quality performance and give a high payoff to some agent misreporting her ability. Similarly, the Information Diffusion Mechanism cannot be applied when considering the execution uncertainty.

\subsection{With Execution Uncertainty}
Consider a more general case with execution uncertainty, each agent herself probably performs the task with different qualities. To incentivize the task information diffusion, we apply the idea of the Information Diffusion Mechanism, i.e., agents are rewarded for their contributions in finding the chosen task performer. However, the Information Diffusion Mechanism fails in incentive compatibility in this setting. To mitigate this issue, we propose the \textbf{Post Execution Verification-based Diffusion Mechanism} (PEV-based Diffusion Mechanism). The PEV-based Diffusion Mechanism chooses the agent to perform the task based on agents' reports. After the task requester verifies the chosen agent's actual performance, the task requester assigns a payoff to the chosen agent based on her actual performance. Given a report profile $\theta' \in \Theta$, define an allocation that maximizes expected social welfare as
\begin{equation*}
\resizebox{.99\linewidth}{!}{$
    \displaystyle
    \pi^*(\theta') = \left\{ \pi_i^*(\theta') = \mathbb{I}\left(i\in \arg\max_{k\in I(\theta')} \{ \mathbb{E}_{f'_k}[Q_k] - c'_k \}\right) \right\}_{i\in N}.
$}
\end{equation*}
The PEV-based Diffusion Mechanism is defined as follows.
\begin{framed}
\noindent\textbf{PEV-based Diffusion Mechanism}
 
\noindent\rule{\textwidth}{0.5pt}
 
\noindent\textsc{Input}: a report profile $\theta'$.
 
\noindent\rule{\textwidth}{0.5pt}
\noindent
1. Choose $j \in \arg\max\nolimits_{i \in N} \{ \mathbb{E}_{f'_i}[Q_i] - c'_i \}$ with random tie breaking.

\noindent
2. Set $\pi_i = 0, p_i = 0$ for all $i \notin I(\theta')$ and all $i\notin ct_j(\theta')$.

\noindent
3. For each agent $i\in ct_j(\theta')$, compute $w_i = \sum_{k \neq i} \pi'_k \left(\mathbb{E}_{f_k'} \left[ Q_k \right] -  c'_k\right)$
    where $\pi' = \pi^*((nil,\theta'_{-i}))$.

\noindent
4. Let the sequence order of agents in $ct_j(\theta')$ be $(s,$ $i_1,$ $i_2,$ $\dots,$ $i_m)$, where $i_{m} = j$. For $k=1:m-1$,
    \[ \pi_{i_k} = \begin{cases} 1 & \text{if } \mathbb{E}_{f_{i_k}'}[Q_{i_k}] - c_{i_k}' = w_{i_{k+1}} \\ & \text{ and } \sum_{l=1}^{k-1} \pi_{i_l} = 0 \\ 0 & \text{otherwise} \end{cases} \]
    and $\pi_j = 1$ if $\sum_{k=1}^{m-1} \pi_{i_k} = 0$.

\noindent
5. Suppose $\pi_{i_t} = 1$ ($1\leq t\leq m$). After agent $i_t$ finishes the task, the task requester receives $q_\pi$.

\noindent
6. The payoff of each agent $i_k\in ct_j(\theta')$ is defined as
    \[ p_{i_k} = \begin{cases} w_{i_{k+1}} - w_{i_{k}} & k < t \\ q_\pi - w_{i_t} & k = t \\ 0 & k > t \end{cases} \]

 
\noindent\rule{\textwidth}{0.5pt}
 
\noindent\textsc{Output}: the allocation $\pi$ and the payoff $p$.
\end{framed}
Intuitively, the PEV-based Diffusion Mechanism allocates the task to one agent in the sequence of critical agents. The chosen agent achieves the highest expected welfare when the next agent in the sequence did not participate. The payoff of the chosen agent is determined by her actual execution quality. In the sequence, each critical agent before the chosen agent gets a payoff based on her contribution. Each critical agent's contribution is measured by the increase of the expected social welfare from her participation when the next critical agent did not participate. We give an example to show how the PEV-based Diffusion Mechanism works.


\begin{example}

Consider the network in Figure~\ref{fig:ex3}, suppose that $Q = \{1,2,\cdots,10\}$, all agents' quality distributions $f$ and costs $c$ in their types are listed in the Table~\ref{tab:types1}. Thus, agent $9$ is the one maximizing expected social welfare and the sequence order of her critical agents is $(s, 2, 6, 9)$. We can compute that $w_2 = \mathbb{E}_{f_3}[Q_3] - c_3 = 4$, $w_6 = \mathbb{E}_{f_3}[Q_3] - c_3 = 4$ and $w_9 = \mathbb{E}_{f_{10}}[Q_{10}] - c_{10} = 4.5$. Applying the PEV-based Diffusion Mechanism, we choose agent $9$ to perform the task.
\begin{figure}[!htbp]
    \begin{minipage}{1\linewidth}
    \centering
    \includegraphics[scale = 0.45]{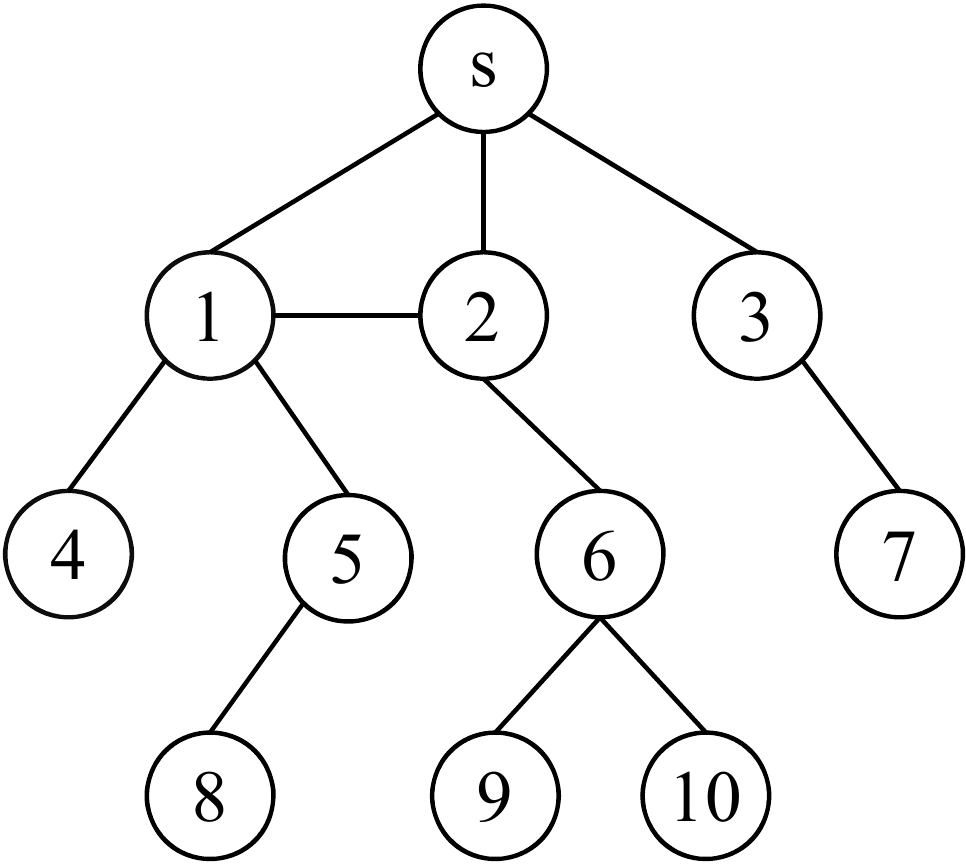}
    \caption{An example for the PEV-based Diffusion Mechanism}
    \label{fig:ex3}
    \end{minipage}
    \begin{minipage}{1\linewidth}
    \centering
    \begin{tabular}{c|c|c}
    \toprule
      $i$   &  $f_i$ & $c_i$ \\
      \midrule
         1 &  \makecell{$f_1(2) = .5, f_1(3) = .5$}  & \makecell{$c_1 = 0.5$} \\
         \midrule
         2 & \makecell{$f_2(1) = 1$} & \makecell{$c_2 = 0.2$} \\
         \midrule
         3 & \makecell{$f_{3}(5) = 1$}  & \makecell{$c_3 = 1$ } \\
         \midrule
         4 & \makecell{$f_4(3) = 1$} & \makecell{$c_4 = 1$ } \\
         \midrule
         5 & \makecell{$f_5(4) = .4, f_5(6) = .6$ } & \makecell{$c_5 = 1.6$} \\
         \midrule
         6 & \makecell{$f_{6}(3) = .3, f_{6}(4) = .6, f_{6}(7) = .1$} & \makecell{$c_6 = 0.9$ } \\
         \midrule
         7 & \makecell{$f_7(6) = .5, f_7(8) = .5$} & \makecell{$c_7 = 4.2$ }\\
         \midrule
         8 & \makecell{$f_8(1) = .2, f_8(3) = .8$ } & \makecell{$c_8 = 0$ } \\
         \midrule
         9 & \makecell{$f_9(8) = .8, f_9(10) = .2$ } & \makecell{ $c_9 = 1$ } \\
         \midrule
         10 & \makecell{$f_{10}(4) = .5, f_{10}(5) = .3, f_{10}(6) = .2$ } & \makecell{ $c_{10} = 0.2$ }  \\ 
    \bottomrule
    \end{tabular}
    \captionof{table}{Types of the agents in Figure~\ref{fig:ex3}.}
    \label{tab:types1}
    \end{minipage}
\end{figure}
Suppose that the actual execution quality of agent $9$
is $q_\pi = 8$. Then, the payoffs given to agents $2$, $6$ and $9$ are $p_2 = w_6 - w_2 = 0$, $p_6 = w_9 - w_6 = 0.5$ and $p_9 = q_{\pi} - w_9 = 3.5$, so their utilities will be $u_2 = p_2 = 0$, $u_6 = p_6 = 0.5$ and $u_9 = p_9 - c_9 = 2.5$. Finally, the utility of the task requester is $u_s = q_\pi - (p_2 + p_6 + p_9) = 4$.
\end{example}



In the following, we will prove the properties of the PEV-based Diffusion Mechanism by first giving two lemmas, and the proofs are shown in Appendix A.1 and A.2.
\begin{lemma}\label{lemma:forir}
Given a report profile $\theta' \in \Theta$, for each agent $j$, the set of her critical agents $ct_j(\theta')$ and the sequence order of critical agents $(s,$ $i_1,$ $i_2,$ $\dots,$ $i_m)$ with $i_m = j$, we have $w_{i_k} \leq w_{i_{k'}}$ for all $k < k'$.
\end{lemma}
\begin{lemma}\label{lemma:ind}
In the PEV-based Diffusion Mechanism, agent $i$'s payoff $p_i$ is independent of $f_i'$ and $c_i'$ for all $i \in N$ and $\theta' \in \Theta$.
\end{lemma}
With these two lemmas, we show that our mechanism satisfies incentive compatibility, individual rationality and weakly budget balance in the following theorem.

\begin{theorem} \label{thm2}
The PEV-based Diffusion Mechanism is individually rational, incentive compatible and weakly budget balanced.
\end{theorem}
\begin{proof}
\noindent \textbf{Individual Rationality.} For each agent $i \in N$, her type $\theta_i$, for all other agents' report profile $\theta_{-i}' \in \Theta_{-i}$, assume that $\pi^*_j((\theta_i,\theta_{-i}')) = 1$. Let $ct_{j}((\theta_i,\theta_{-i}'))$ be the set of agent $j$'s critical agents, and $(s, i_1, i_2, $ $  \dots, i_m)$ be the sequence order of agents in $ct_{j}((\theta_i,\theta_{-i}'))$. Applying the PEV-based Diffusion Mechanism, let $\pi$ be the allocation and $i_t$ be the selected agent, $1 \leq t \leq m$. For all agents who are not in $I((\theta_i,\theta_{-i}'))$, their payoffs and costs are zero, which means that their utilities are all zero. For each agent $i \in I(\theta')$, her payoff depends on the relationship between her and agent $i_t$. 

\textbf{(i)} If agent $i$ is not agent $i_t$'s critical agent, i.e., $i \notin ct_{j}$ or $i = i_k \in ct_j$ with $k > t$, her payoff and actual cost are zero.

\textbf{(ii)} If $i\neq i_t$, but $i$ is $i_t$'s critical agent, i.e., $i = i_k \in ct_j$ with $k < t$, her utility is $u_{i_k} = p_{i_k} = w_{i_{k+1}} - w_{i_k}$. According to Lemma~\ref{lemma:forir}, her expected utility is always non-negative.

\textbf{(iii)} If agent $i$ is the chosen agent $i_t$, her expected utility is $\mathbb{E}_{f_i}\left[ q_{\pi} - w_{i_t} -c_{i_t} \right] 
    =\ \mathbb{E}_{f_i}\left[ q_{\pi}  - c_{i_t} \right]  -  \mathbb{E}_{f_{\pi'}'}[ q_{\pi'} - \sum\nolimits_{k\neq i} \pi'_k c'_k ] \geq 0$,
where $\pi' = \pi^* ((nil,\theta_{-i}'))$. The inequality holds since agent $i$ must be the one maximizing the expected social welfare for agents in $I((nil,\theta_{-i}')) \cup \{i\}$,
otherwise she won't be selected.

Hence, each agent's expected utility is non-negative when truthfully reporting, and our mechanism is \textbf{IR}.

\noindent \textbf{Incentive Compatibility.} Given agent $i$'s type $\theta_i$, let $\theta_{-i}'$ be all other agents' report profile. 
Assume agent $j$ is the agent maximizing the expected social welfare when $i$ truthfully reports, i.e., $\pi^*_j((\theta_i,\theta_{-i}')) = 1$. Let $(s, i_1, i_2, \dots, i_m)$ be the sequence order of agents in $ct_{j}((\theta_i,\theta_{-i}'))$. Applying the PEV-based Diffusion Mechanism, let $i_t$ be the selected agent under $(\theta_i,\theta_{-i}')$, $1 \leq t \leq m$. Note that for all $i \notin I((\theta_i,\theta_{-i}'))$, agent $i$'s expected utility is zero for her all possible reports. Thus, we only consider the cases where agent $i$ is in $I((\theta_i,\theta_{-i}'))$. 


\textbf{(i)} If $i \notin ct_j((\theta_i,\theta_{-i}'))$, $i$ cannot change the allocation by only misreporting $r_i'$, and then $i$'s expected utility is zero. Her expected utility changes only when she misreports $f_i'$ or $c_i'$ to be selected. Now her expected utility is $\mathbb{E}_{f_i} \left[ q - w_{i} - c_i \right]$, where $q$ is her actual completion quality and $w_i = $ $ \mathbb{E}_{f_{\pi'}'}\left[ q_{\pi'} \right] - \sum_{k\neq i} \pi'_k c'_k $ and $\pi' = \pi^*((nil,\theta'_{-i}))$. $\mathbb{E}_{f_i} \left[ Q_{i} - c_i \right] \leq \mathbb{E}_{f_j'} \left[ Q_j - c_j \right]$ holds for $\pi^*_j((\theta_i,\theta_{-i}')) = 1$. Since $j \in I((nil,\theta'_{-i}))$, we have $\pi_j' = 1$ and $w_i = \mathbb{E}_{f_j'} \left[ Q_j - c_j' \right]$. Then, her utility is $\mathbb{E}_{f_i} \left[ q - w_{i} - c_i \right] =\  \mathbb{E}_{f_i} \left[ Q_{i} - c_i \right] - \mathbb{E}_{f_j'} \left[ Q_j - c_j' \right] \leq 0$.
Then, agent $i$'s utility will not increase if she misreports.

\textbf{(ii)} If $i = i_k \in ct_j((\theta_i,\theta_{-i}'))$ with $k>t$, agent $i_t$ is still selected for all possible $\theta_i'$ and agent $i$'s expected utility is always zero.

\textbf{(iii)} If $i = i_k\in ct_j((\theta_i,\theta_{-i}'))$ with $k<t$, agent $i$'s expected utility is $\mathbb{E}_{f_i}\left[w_{i_{k+1}} - w_{i_{k}}\right] $ $ = w_{i_{k+1}} - w_{i_{k}}$, where $w_{l}$ is the maximum expected social welfare for all agents in $I((nil,\hat{\theta}_{-l}))$ with $\hat{\theta} = (\theta_i,\theta_{-i}')$.
If $i$ misreports her type as $\theta_i'$, denote all agents' report profile by $\hat{\theta}' = (\theta_i',\theta_{-i}')$. 
Applying the PEV-based Diffusion Mechanism, let $j'$ be the selected agent under $\hat{\theta}'$, and $(s, i'_1, i'_2, \dots, i'_{m'})$ be the sequence order of agents in $ct_{j'}(\hat{\theta}')$.
\begin{enumerate}
    \item If $i \notin ct_{j'}(\hat{\theta}')$, her expected utility decreases to zero.
    \item If agent $i$ is a critical agent of agent $j'$, agent $j'$ is also a critical agent of $j$, i.e., $j' = i_{k'} \in ct_j(\hat{\theta})$ with $k' > k$, agent $i$'s expected utility is unchanged.
    \item If agent $i$ is a critical agent of $j'$ and $j'$ is not critical agent of agent $j$, i.e., $i = i_{k'}' \in ct_{j'}(\hat{\theta}')$ and $j' \notin ct_{j}(\hat{\theta})$, agent $i$'s expected utility is $\mathbb{E}_{f_i} [w_{i_{k'+1}'}' - w_{i'_{k'}}' ] = w_{i_{k'+1}'}' - w_{i'_{k'}}'$, where $w_l'$ is the maximum expected social welfare for agents in $I((nil,\hat{\theta}_{-l}'))$. $w_{i'_{k'}}' = w_{i_{k}}$ holds since $I((nil,\hat{\theta}_{-i}')) = I((nil,\hat{\theta}_{-i}))$. Since $j' \notin ct_{j}(\hat{\theta}')$, agent $i_{k+1}$ cannot be informed of the task information. Thus, $I((nil,\hat{\theta}'_{-i'_{k'+1}}))$ $\subset$ $ I((nil,\hat{\theta}_{-i_{k+1}}))$ and $w_{i'_{k'+1}}' \leq w_{i_{k+1}}$. Accordingly, we have $w_{i'_{k'+1}}' - w_{i'_{k'}}' \leq w_{i_{k+1}} - w_{i_{k}}$, which means $i$'s expected utility decreases.
    
    \item If $i = j'$, let $q$ be $i$'s completion quality. Her expected utility is $\mathbb{E}_{f_i} \left[q - w_{i_k} - c_i \right] = \mathbb{E}_{f_i} \left[q - c_i \right] - w_{i_k}$.
    Since $i \in I((nil,\hat{\theta}_{-i_{k+1}}))$, and agent $i$ is not the one maximizing the expected social welfare when without agent $i_{k+1}$'s participation under $\hat{\theta}$, we have that $\mathbb{E}_{f_i} \left[q - c_i \right] - w_{i_k} - (w_{i_{k+1}} - w_{i_k})
       =\ \mathbb{E}_{f_i} \left[q - c_i \right] - w_{i_{k+1}}
       =\  \mathbb{E}_{f_i} \left[Q_i - c_i \right] - w_{i_{k+1}}\leq 0$, which means that agent $i$'s expected utility decreases when misreporting.
\end{enumerate}

\textbf{(iv)} If $i = i_t$, her expected utility is $\mathbb{E}_{f_i} \left[q_{\pi} - c_i - w_{i_t} \right]$.
If agent $i$ misreports her type as $\theta_i'$, denote all agents' report profile by $\hat{\theta}' = (\theta_i',\theta_{-i}')$ and the original one by $\hat{\theta} = (\theta_i,\theta_{-i}')$. 
Applying the PEV-based Diffusion Mechanism, let $j'$ be the selected agent under $\hat{\theta}'$ and $ct_{j'}(\hat{\theta}')= (s, i'_1, i'_2, \dots, i'_{m'})$.
\begin{enumerate}
    \item If $i = j'$, her expected utility is unchanged, which can be referred from the Lemma~\ref{lemma:ind}.
    \item If $i \notin ct_{j'}(\hat{\theta}')$, her expected utility decreases to zero.
    \item If $i\neq j'$, but agent $i$ is one critical agent of agent $j'$, i.e, $i = i_k' \in ct_{j'}(\hat{\theta}')$ and $k' < m'$, her expected utility is $\mathbb{E}_{f_i} [w_{i_{k+1}'}' - w_{i_{k}'}' ] = w_{i_{k+1}'}' - w_{i_{k}'}'$, where $w_l'$ is the maximum expected social welfare under $(nil,\hat{\theta}_{-l}')$. Then, $w_{i'_{k}}' = w_{i_{t}}$ holds for $I((nil,\hat{\theta}_{-i}')) = I((nil,\hat{\theta}_{-i}))$. 
    
    Since $\pi_i(\hat{\theta})=1$, then we have $i = j$ or $ w_{i_{t+1}}=\mathbb{E}_{f_i}\left[Q_i - c_i \right]$ if $t<m$. In the first case, $\mathbb{E}_{f_i} \left[q_{\pi} - c_i \right] \geq w_{i_{k+1}'}'$ always holds. In the second case, if $j' \in ct_j(\hat{\theta})$, we have $i_{k+1}' = i_{t+1}$, which means that $w_{i_{k+1}'}' \leq w_{i_{t+1}} = \mathbb{E}_{f_i}\left[Q_i - c_i \right]$. If $j' \notin ct_j(\hat{\theta})$, agent $i_{t+1}$ must not be informed of the task information. Then, we have $I((nil,\hat{\theta}_{i_{k+1}'}')) \subset I((nil,\hat{\theta}_{i_{t+1}}))$ and $w_{i_{k+1}'}' \leq w_{i_{t+1}} = \mathbb{E}_{f_i}\left[Q_i - c_i \right]$. Thus, the relationship between her current utility and her original one is $w_{i_{k+1}'} - w_{i_{k}'} - (\mathbb{E}_{f_i} \left[q_{\pi} - c_i \right] - w_{i_t})
    =\  w_{i_{k+1}'}  - \mathbb{E}_{f_i} \left[q_{\pi} - c_i \right] =\  w_{i_{k+1}'}  - \mathbb{E}_{f_i} \left[Q_i - c_i \right]\leq 0 $.


\end{enumerate}

In summary, for each $i \in N$, agent $i$'s expected utility is maximized when she is truthful. Hence, the PEV-based Diffusion Mechanism is \textbf{IC}.

\noindent \textbf{Weakly Budget Balance.} Given a report profile $\theta' \in \Theta$, let $\pi$ be the allocation $\pi$ under $\theta'$, the expected utility of $s$ is
\begin{align*}
    &\mathbb{E}_{f_\pi'} \left[ q_\pi - \sum\nolimits_{i\in N} p_i(\theta', q_{\pi}) \right] \\
    =\ & \mathbb{E}_{f_\pi'} \left[ q_\pi - \left( q_\pi - w_{i_t} \right) - \sum\nolimits_{k=1}^{t-1}  \left(w_{i_{k+1}} - w_{i_{k}} \right) \right] \\
    =\ & \mathbb{E}_{f_\pi'} \left[ w_{i_1} \right] = w_{i_1} \geq 0
\end{align*}
Hence, the PEV-based Diffusion Mechanism is \textbf{WBB}.
\end{proof}

Notice that the expected utility of the task requester equals $w_{i_1}$, i.e., the maximum expected social welfare when the first agent in the critical agent sequence did not participate. It also means that the expected utility of the task requester will be no less than that when only the neighbours of the task requester participate. Notice that our mechanism cannot always guarantee efficiency since the task may not be allocated to the best worker. Below, we give a proposition to show that no mechanism simultaneously satisfies IR, IC, WBB and efficiency. The proof is shown in Appendix A.3.

\begin{proposition}
There is no verified contract mechanism satisfying individual rationality, incentive compatibility, weakly budget balanced and efficiency in the social networks.
\end{proposition}

\section{Conclusion}\label{sec:conclude}
In this paper, we focus on the single-task allocation problem in the social network. We propose the PEV-based Diffusion Mechanism to incentivize agents to propagate the task information, thus involving more participants and finding a better worker.
Then, the PEV-based Diffusion Mechanism can increase the expected utility of the task requester. There also exist open problems worth further investigations. For example, applicable mechanisms are still missing in multiple-task allocation settings where tasks have combinatorial qualities or there exist dependencies between tasks. More importantly, we do expect our work can inspire the research community in the future.



\bibliographystyle{named}
\bibliography{ijcai22}

\newpage
\appendix
\section{Appendix}
\setcounter{lemma}{0}
\renewcommand{\thelemma}{\arabic{lemma}}
\subsection{Proof of Lemma 1}
\label{lemma1_proof}
\begin{lemma}
Given a report profile $\theta' \in \Theta$, for each agent $j$, the set of her critical agents $ct_j(\theta')$ and the sequence order of critical agents $(s,$ $i_1,$ $i_2,$ $\dots,$ $i_m)$ with $i_m = j$, we have $w_{i_k} \leq w_{i_{k'}}$ for all $k < k'$.
\end{lemma}
\begin{proof}
With the above setting, $i_k \in ct_{i_{k'}}(\theta')$ holds for all $k < k'$. $w_{i_k}$ and $w_{i_{k'}}$ are the maximum expected social welfare for agents in $I((nil,\theta'_{-i_k}))$ and $I((nil,\theta'_{-i_{k'}}))$ respectively. Since agent $i_k$ is before agent $i_k'$ in the sequence, we have that $I((nil,\theta'_{-i_k})) \subset I((nil,\theta'_{-i_{k'}}))$. Then, $w_{i_k} \leq w_{i_{k'}}$ holds for all $k < k'$.
\end{proof}
\subsection{Proof of Lemma 2}\label{lemma2_proof}
\begin{lemma}
In the PEV-based Diffusion Mechanism, agent $i$'s payoff $p_i$ is independent of $f_i'$ and $c_i'$ for all $i \in N$ and $\theta' \in \Theta$.
\end{lemma}
\begin{proof}
Given a report profile $\theta'$, let $\pi$ be the allocation in the PEV-based Diffusion Mechanism. Assume that $\pi^*_j(\theta') $ $= 1$, $ct_j(\theta') = (s, i_1,i_2,\dots,i_m)$ and $i_t$ is the selected agent. For each agent $i \notin I(\theta')$, her payoff is zero. For each other agent $i \in I(\theta')$, we need to consider several cases.

\begin{itemize}
    \item If agent $i$ is not a critical agent of $i_t$, i.e., $i \notin ct_j (\theta')$ or $i = i_k \in ct_j(\theta')$ with $k >t$, her payoff is zero.
    \item If agent $i$ is not the selected agent, but she is $i_t$'s critical agent, i.e., $i = i_k \in ct_j (\theta')$ with $k < t$, her payoff is $w_{i_{k+1}} - w_{i_k}$. Agent $i$ is not the one maximizing the expected social welfare when without agent $i_{k+1}$'s participation. Thus, $w_{i_{k+1}}$ and $w_{i_k}$ are both independent of $f_i'$ and $c_i'$, her payoff is independent of $f_i'$ and $c_i'$.
    \item If agent $i$ is the selected agent, her payoff is $q_\pi - w_{i_t}$. Her execution quality $q_\pi$ is only determined by $f_i$ in her type and $w_{i_t}$ is independent of $f_i'$ and $c_i'$.
\end{itemize}

Hence, for all agent $i \in N$, $i$'s payoff is independent of $f_i'$ and $c_i'$ in the PEV-based Diffusion Mechanism.
\end{proof}

\subsection{Proof of Proposition 1}
\setcounter{proposition}{0}
\renewcommand{\theproposition}{\arabic{proposition}}
\begin{proposition}
There is no verified contract mechanism satisfying individual rationality, incentive compatibility, weakly budget balanced and efficiency in the social networks.
\end{proposition}
\begin{proof}
We can prove this statement by showing that no individually rational, incentive compatible and efficient verified contract mechanism can satisfy weakly budget balance in the social networks. 
\begin{figure}[!h]
    \centering
    \includegraphics[scale=0.5]{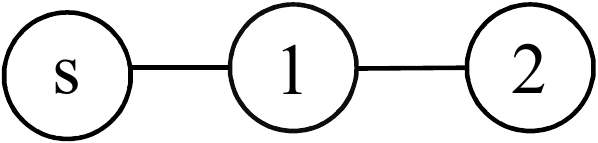}
    \caption{A counterexample: the task requester $s$ is connected to agent $1$, and agent $1$ has two neighbours: $s$ and agent $2$.}
    \label{fig:4}
\end{figure}

Let $\mathcal{M}$ be an IR, IC and efficient verified contract mechanism.
Consider a graph shown in figure~\ref{fig:4}, there are only three agents. Given the task to be performed, and suppose that the performance level set is $Q$. Assume that $f_1(q_1)=1$, $f_2(q_2) = 1$, $q_1=q_2$ and $\mathbb{E}_{f_{2}}\left[Q_2 - c_2\right] > \mathbb{E}_{f_{1}}\left[Q_1 - c_1\right] > 0$.
\begin{itemize}
    \item When agent $1$ does not invite agent $2$, the mechanism $\mathcal{M}$ will allocate the task to agent $1$ and agent $1$ must get a payoff of $q_1$. Otherwise, agent $1$ can misreport her cost $c_1'$.
    \item When agent $1$ invites agent $2$, agent $2$'s payoff should be at least $c_2'$. However, to guarantee IC, agent $1$ should get a payoff of at least $q_1-c_1'$. Then, agent $1$ must get at least $q_1-c_2$ otherwise agent $1$ can misreport her cost to get higher payoff. Similarly, agent $2$'s payoff must be at least $c_1$. Then, the utility of the sponsor will be at most $q_2 - p_1 -p_2 \leq q_2-(q_1-c_2)-c_1 = c_2-c_1<0$. 
    Then, the mechanism $\mathcal{M}$ is not WBB.
\end{itemize}
\end{proof}

\end{document}